# Electrochemical reaction in single layer $MoS_2$: nanopores opened atom by atom

J. Feng[1], K. Liu[1], M.Graf[1], M.Lihter[1,2], R. D. Bulushev[1], D. Dumcenco[3], D.T.L. Alexander[4], D. Krasnozhon[3], T. Vuletic[2], A.Kis[3], and A.Radenovic[1]

[1]*Laboratory of Nanoscale Biology, Institute of Bioengineering, School of Engineering, EPFL, 1015 Lausanne, Switzerland*
[2]*Institut za fiziku, Bijenička 46,Zagreb, Croatia*
[3] *Laboratory of Nanoscale Electronics and Structure, Institute of Electrical Engineering , School of Engineering, EPFL, 1015 Lausanne, Switzerland*
[4] *Centre Interdisciplinaire de Microscopie Électronique (CIME), EPFL, 1015 Lausanne, Switzerland*

**Correspondence should be addressed* ke.liu@epfl.ch *and* aleksandra.radenovic@epfl.ch

## ABSTRACT

**Ultrathin nanopore membranes based on 2D materials have demonstrated ultimate resolution toward DNA sequencing. Among them, molybdenum disulphide ($MoS_2$) shows long-term stability as well as superior sensitivity enabling high throughput performance. The traditional method of fabricating nanopores with nanometer precision is based on the use of focused electron beams in transmission electron microscope (TEM). This nanopore fabrication process is time-consuming, expensive, not scalable and hard to control below 1 nm. Here, we exploited the electrochemical activity of $MoS_2$ and developed a convenient and scalable method to controllably make nanopores in single-layer $MoS_2$ with sub-nanometer precision using electrochemical reaction (ECR). The electrochemical reaction on the surface of single-layer $MoS_2$ is initiated at the**

**location of defects or single atom vacancy, followed by the successive removals of individual atoms or unit cells from single-layer $MoS_2$ lattice and finally formation of a nanopore. Step-like features in the ionic current through the growing nanopore provide direct feedback on the nanopore size inferred from a widely used conductance vs. pore size model. Furthermore, DNA translocations can be detected in-situ when as-fabricated $MoS_2$ nanopores are used. The atomic resolution and accessibility of this approach paves the way for mass production of nanopores in 2D membranes for potential solid-state nanopore sequencing.**

**Keywords: solid-state nanopores, 2D materials, molybdenum disulphide ($MoS_2$), electrochemical reaction (ECR), DNA translocation**

Fabrication of nanostructures with sub-nanometer, or even single-atom precision has been a long-term goal for nanotechnology. The rise of graphene[1] and recently other 2D materials, such as the single-layer molybdenum disulphide ($MoS_2$)[2], offers an ideal platform for such a purpose, due to their highly ordered lattice in two dimensions. Fabrication of solid-state nanopores that are used in single-molecule sensing[3, 4] would benefit tremendously from such a nanoscale fabrication method. Conceptually, nanopore sensing is based on a single, nanometer sized aperture located on a nanometer thin membrane; by monitoring the changes in the ionic current it is possible to detect electrophoretically driven biomolecular translocations in a high throughput manner, while revealing localized information of the analyte. Although conceptually simple, the method is still limited to laboratory use[5] since the formation of a single solid-state nanopore with sub-nm precision relies heavily on high-end instrumentation, such as a transmission electron microscope (TEM)[5] and a well-trained TEM user. This nanopore fabrication process is time-consuming, expensive, not scalable and hard to control below 1 nm. Many efforts, such as chemical wet-etching of silicon[6] or polyethylene terephthalate film[7] have been carried out towards mass production of nanopores. Recently, a pioneering and simple method has been reported using controlled dielectric breakdown to make individual nanopores (3-30 nm diameter) on insulating silicon nitride membranes (5-30 nm thick) without the need of TEM[8, 9] .

Among solid-state pores, the highest single to noise ratio (SNR) and sensitivity has been reported for the atomically thin nanopore membranes made from 2D materials, such as graphene[10-12], boron nitride[13] and $MoS_2$[14]. Theoretically, base by base recognition can be achieved since membrane thicknesses have comparable values with the base-stacking distance (0.34 nm). Therefore they hold promise for the so-called 3rd generation DNA sequencers. Recently, we have demonstrated the first realization of single nucleotide identification in small $MoS_2$ nanopores (< 4 nm), where we introduced a viscosity gradient system based on room temperature ionic liquids (RTILs) to slow down DNA translocation[15]. The differentiation of nucleotides is based on their ionic current signal and relies strongly on the pore diameter[15]. A controllable nanopore fabrication method, which allows mass production of $MoS_2$ nanopores below 4 nm with atomic precision, is therefore highly desired.

$MoS_2$, as a member of transition metal dichalcogenide (TMD) family, has rich electrochemical properties such as catalytic hydrogen generation[16]. During the past several decades, scanning probe microscopes (SPMs) such as scanning tunneling microscopes (STM) and atomic force microscopes (AFM), demonstrated ability to craft nanostructures with an atom/molecule resolution. In SPM, using tip-induced electrochemical reaction, it is possible to engineer nanostructures or make holes in layered TMDs ($WSe_2$, $SnSe_2$, $MoSe_2$ or $MoS_2$). The mechanism can be understood as a surface electrochemical reaction scheme induced via the electric field generated by the SPM tip[17, 18]. The oxidation process starts preferably at the surface defects when the voltage threshold (1.2 V in case of $WSe_2$)[19] for oxidation is reached and allows variety of nanoengineering means. However, it is still challenging to make nanopores on suspended membranes using SPMs, while on the other hand implementation of SPMs instrument in nanopore fabrication is comparable to TEMs in terms of cost and complexity.

Here we present *in-sit*u application of the electrochemical reaction (ECR) for fabrication of individual nanopores on single-layer $MoS_2$, with the electric field generated by Ag/AgCl electrodes away from the membrane. ECR starts for a certain critical voltage bias at a defect/vacancy present in the $MoS_2$ membrane.

Importantly, in the course of ECR fabrication we observe – and we are able to control - the successive removal of single or few $MoS_2$ units from the monolayer $MoS_2$ membranes. In this way we accomplish the atom-by-atom nanopore engineering. To the best of our knowledge, this is the first example of nanopore engineering on single-layer $MoS_2$ membranes with atomic precision utilizing ECR.

The procedure for fabricating $MoS_2$ nanopores using ECR is schematically illustrated in **Fig. 1a**, where two chambers (*cis* and *trans*) are filled with aqueous buffer (1M KCl, pH 7.4) and biased by a pair of Ag/AgCl electrodes which are separated by a single-layer $MoS_2$ membrane. Presence of an active site such as single-atom vacancy[19] facilitates the removal of individual atoms and $MoS_2$ unit cells from $MoS_2$ lattice by ECR at voltages higher than the oxidation potential of $MoS_2$ in aqueous media. This process is facilitated by the electric field focusing by the pore itself. To form freestanding membranes, CVD-grown monolayer $MoS_2$[20] transferred from a sapphire substrate is suspended over focused ion beam (FIB) defined openings that ranged from 80 nm to 300 nm in diameter and were centered in a 20 nm thick $SiN_x$ membrane (see **Fig. 1b)**. A typical optical image of the transferred triangular flake of CVD-grown monolayer $MoS_2$ on the supporting silicon nitride membrane is shown in **Fig. 1c**. The freestanding $MoS_2$ membrane above the FIB defined opening can be further identified under TEM with low magnification (5 k×) as shown in **Fig. 1d**. $MoS_2$ flake is further characterized by Energy-dispersive X-ray spectroscopy (EDX) in TEM to reveal the chemical composition. Elements of Mo and S are abundant in the triangular areas as shown in **SI Fig. 1**.When moving to the high magnification (1 M ×) and focusing on the freestanding portion of $MoS_2$ over the FIB opening, the atomic structure of $MoS_2$ can be clearly resolved as shown in **Fig. 1e**, and the diffractogram reflects the hexagonal symmetry of $MoS_2$, as shown in the inset of **Fig. 1e.**

When an intact $MoS_2$ membrane is mounted into a custom made microfluidic flow-cell filled with an aqueous buffer, transmembrane potential is applied using a pair of Ag/AgCl electrodes. For a voltage bias below the potential for electrochemical oxidation, small leakage current is normally detected, typically on the order from tens to hundreds of picoamperes depending on the number of defects in the 2D membrane[21]. As shown in the **SI Fig. 2**, the leakage current displays a non-ohmic characteristic. To reach the critical voltage bias value for ECR, the potential is gradually stepped, as shown in **Fig. 2a.** When the applied voltage is stepped up to 0.8 V (a critical voltage, indicated by the arrow), an increase of baseline current immediately occurs. This time-point indicates the nanopore creation which is associated to the electrochemical dissolution of $MoS_2$ enhanced by the ion flow focused on the active site as shown (**SI Fig. 3**).

In contrast to the avalanche-like dielectric breakdown process in silicon nitride, where a typically 10-minute waiting time for the filling of charge traps[9] under the application of critical voltage (> 10 V) is needed before breakdown occurs, electrochemical dissolution happens spontaneously at the critical voltage.

In addition, the observed rise of ionic current shows a quite slow rate (~ 0.4 nA/s). The control on the nanopore size is obtained by using an automatic feedback to cut off the voltage once the desired current/conductance threshold is reached. This feedback also helps to avoid multiple pore formation. Owing to the limited rates of electrochemical reaction, the $MoS_2$ nanopore sculpting process is quite slow, occurring on time scales of dozens of seconds to several minutes. **Fig. 2 a** gives an example of ionic current trace to reach the threshold of 20 nA, for the critical voltage of 0.8 V.

Taking the advantage of existing theoretical insights to model the conductance-pore size relation,[22] the conductance of the nanopore (*G*) can be described by[22]

$$G = \sigma \left[ \frac{4L}{\pi d^2} + \frac{1}{d} \right]^{-1} \quad (1)$$

where *σ, L* and *d* are the ionic conductivity of solution, membrane thickness and nanopore diameter, respectively. Using this relation in combination with feedback on ECR that immediately stops the voltage once the desired pore conductance – that corresponds to a certain pore size - is reached, we were able to fabricate pores ranging in diameter from 1-20 nm. **Fig. 2b** reveals current-voltage (I-V) characteristics of $MoS_2$ nanopores fabricated by ECR with different estimated sizes ranging from 1 nm to 20 nm. The symmetric and linear I-V curves also imply the well-defined shape of the fabricated pores. Similarly, as shown in the inset of the **Fig. 2b,** I-V characteristics across the membrane have been investigated in situ before and after ECR, confirming the pore formation.

To further verify the size of fabricated $MoS_2$ nanopores, TEM has been used to image the newly formed nanopore. Exposure of 2D materials to electron radiation can induce large area damage and also open pores, as reported for both graphene[23-25] and $MoS_2$[26]. To minimize this risk we imaged the pore using Cs-corrected high-resolution TEM (Cs-TEM) at a primary beam energy of 80 keV, using a double-corrected FEI Titan Themis 60-300. (We note that, while Cs-corrected scanning TEM (Cs-STEM) gives more directly-interpretable atomic structure contrast, its application here was precluded because of residual hydrocarbon contamination from the prior ECR process condensing rapidly under the Å-sized probe during imaging. A better sample cleaning procedure would be required to realize successfully Cs-STEM imaging of the ECR pore.) We first aligned the imaging condition on the unsuspended portion of $MoS_2$ outside of the FIB opening and then quickly scanned the suspended monolayer region to find and image the ECR-fabricated pore, all the while taking care to irradiate it minimally **Fig, 2c** shows the resultant image of an ECR-fabricated $MoS_2$ nanopore; its current voltage characteristics taken after ECR are shown in **SI Fig 4a**, together with an Cs-TEM imaging overview of the surrounding region **SI Fig 4b**.

The reliability of fabricating $MoS_2$ nanopores using the ECR technique is 90%. A few graphene membranes have also been tested by this method and higher voltages (2-3V) are required to fabricate pores as presented in Supporting Information, with the typical ionic current trace is displayed in **SI Fig 5**.

The described ECR-based pore formation method benefits from the unique crystal structure of transition metal dichalcogenide ($MX_2$) where atoms are situated in tree planes and linked by metal-chalcogenide bonds while in the case of graphene, carbon atoms are in the same plane and 3 bonds need to be removed to release one carbon atom. In addition, to remove carbon atoms, graphene needs to be oxidized to a higher valence state which presumably requires a higher voltage bias.

Despite different chemical compositions of transition metal dichalcogenides ($MX_2$), the pore formation mechanism is in general governed by the electrochemical oxidation reaction that occurs at the location of the defect and requires comparable field strengths to those encountered in SPMs[17, 18]. In our case, mechanical avulsion is highly unlikely to occur since the force is insufficient, similarly to measured results by SPM experiments. The critical voltage 1.2 V for $WSe_2$ is in good agreement with our observations (0.8 V for $MoS_2$), especially if we consider the position in energy of the surface band edges. The physics of the electrochemically fabricated nanopores is determined by the focused electrical field and surface chemistries. The electric field concentrates at surface irregularities or defects which can be considered as surface active sites, and focuses current flow at the site of the pore, and thus locally enhances the electrochemical dissolution, as shown in **Fig 2d.** The surface dissolution chemistries can be understood as a surface bound oxidation scheme with hole capture and electron injection to produce the $MoS_2$ oxidation state[27] as shown in

$$MoS_2 + 11H_2O = MoO_3 + 2SO_4^{2-} + 22H^+ + 18e^- \quad (2)$$

where $MoS_2$ is oxidized into $MoO_3$ which and detached into the solution. We believe this reaction is highly likely to happen considering the electrical potential (voltage bias) range we work with. Due to the current technical limitations of electron energy loss spectroscopy (EELS) analysis in the nanopore vicinity, we cannot exclude the possibility that $MoS_2$ is oxidized to other valence states. Once an active site is removed by the process described above and very small nanopore formed **Fig2 d.**, due to the fact that the nanopore has a much larger resistance than the electrolyte solution, the flux of the ions will converge towards the pore (**SI Fig3**). This focused ionic current through the pore will locally enhance the electrochemical dissolution as previously described by Beale model[28]. In addition, it is possible that the high number of dangling bonds within the nanopore contributes to the more favorable enlargement of a single nanopore rather than nucleation of the many pores. Of course in the presence of many defects, correlated to to the material quality in the suspended area, it is hard to eliminate the possibility that one has created multiple pores. By applying a bias voltage higher than the critical voltage at the beginning of the fabrication process it might be possible to observe the formation of multiple pores. Given the stochastic nature of the pore creation process, with our configuration of voltage steps, multiple simultaneous nanoscale ECR events are highly unlikely. Furthermore, feedback control on the applied voltage to obtain the desirable conductance ensures the formation of a single nanopore. Finally, the formation of a single nanopore is verified by TEM imaging. By establishing the correspondence of nanopore conductance to TEM images of their size, in the future we hope that this step could be be omitted.

The power of ECR-based nanopore fabrication technique, apart from the advantages of being a fast and cheap production lies in the possibility of fine-tuning the diameter of nanopores with unprecedented, single-atom precision. The low nanopore enlarging speed is due to low voltages and the electrochemical dissolution nature of the process. **Fig. 3b** is a 25-second long, continuous pore conductance trace that shows atomic precision during nanopore sculpting process. The trace starts from the critical point indicated in **Fig. 2a.** Fitting to the conductance-nanopore size relation, we can estimate a pore diameter growth rate of about 1 Å per second. After 25 seconds a pore with a diameter 1.9* nm (area of 2,9 $nm^2$) has been formed. The area of such a pore is equivalent to almost exactly N = 34 unit cells of $MoS_2$ where the area of the unit cell $u$ = 0.0864 $nm^2$ (**Fig. 3a**).

To our surprise, the growth curve is not linear but step-like, as shown in **Fig. 3b**. Necessarily, the effective size of the pore enlarges with the same step-like characteristic. To gain insights into these step-like features, we plotted the histogram of current values from this trace in **Fig. 3c**, where 21 individual peaks can be extracted from the histogram.

The sequence of the pore size enlargement steps may be normalized by the unit cell area $u$ and a sequence of $MoS_2$ formula units and Mo and S atoms cleaved (corresponding to 21 current steps) to form the pore may be inferred, as presented in the **Fig. 3c** (for details see Supporting information). Several snapshots of the proposed pore formation process, taken at steps 1,7,14 and 21, are displayed in the **Fig. 3 d**. The full sequence of the pore formation is displayed in ***Supporting Movie 1***. The area of polygons corresponding to the cleaved atom groups follows the honeycomb structure of single-layer $MoS_2$, as presented schematically in **Fig. 3a** and in the cleavage steps superimposed on a Cs-STEM image of $MoS_2$ lattice, **Fig 3 d.**

Here presented, step-like features are commonly observed when working with low voltages ranging from 0.8-2 V. The reproduced step-like features from another devices is shown in **SI Fig. 7**.The observed atomic steps here reveal the ultimate precision (single atoms) that can be reached in engineering nanostructures.

To test the performance of ECR-fabricated pores, we performed DNA translocation experiments and detected the translocation events by the current drops below the baseline current. ECR fabricated $MoS_2$ nanopores consistently produces low-1/*f* noise on the current baseline, which is slightly higher than TEM drilled $MoS_2$ nanopores **(SI Fig.8)**. The major contribution to the 1/f noise in 2D membrane nanopores[29] can be attributed to mechanical fluctuations of the thin membranes. Higher frequency fluctuations are produced by the method itself. Fluctuation noise can be significantly reduced by using a smaller supporting opening,[30] or operating at low temperatures. To show the ability of ECR fabricated nanopore for DNA detection, 2.7 kbp pNEB plasmid DNA is translocated through a relatively large $MoS_2$ nanopore (30-40 nm) to eliminate the pore-DNA interaction and multiple conformation issues. **Fig. 4a**. displays only one-level events indicating an extended (unfolded) DNA conformation, with SNR >10. Scatter plots are used to describe the statistics of DNA translocation as shown in **Fig. 4b.** The signal amplitude also increases linearly with the applied voltage, which is 0.5 nA for 450 mV and 0.38 nA for 300 mV as shown in the histogram **Fig. 4b.** Dwell times are also comparable with DNA translocation through a TEM-drilled $MoS_2$ nanopore of a similar diameter, for the same DNA and under same bias conditions. In addition, λ-DNA (48 k bp) is also translocated through an ECR-fabricated nanopore shown in (**SI Fig. 9**). A noticeable advantage for this nanopore fabrication method is that DNA translocations can be performed *in situ* after ECR and size-control allows on-demand adaptation of the pore size, allowing sizing for the different types of biomolecules, e.g. proteins[31] or DNA-protein complexes[32]. In addition, to verify the single pore formation for the small nanopore sizes <5nm, λ-DNA (48 k bp) is also translocated through 4.3 nm large ECR-fabricated nanopore (**SI Fig. 9**). As shown in **SI Fig 10. a**, obtained from the simplistic model that assumes two pores, conductance drop will strongly depend on the ratio of the two pore sizes. The experimentally observed blockage 11% (**SI, Figure 9**) is in a good agreement with the assumption of single 4.3nm pore. For the larger pore sizes 15-30 nm this simplistic analytical model is less reliable since the conductance drop caused by DNA translocation varies slightly (see Supporting Information).

Apart from nanopore sensors, other applications can be further explored based on the conductance quantification such as selective ion transport, nanoionics[33], and atomic switches or as platforms for understanding electrochemical kinetics[34]. To conclude, we present the atomically controlled electrochemical etching of single-layer $MoS_2$ which we employ to engineer nanopores with sub-nanometer precision. The fabricated $MoS_2$ nanopores are carefully characterized by I-V characteristics and their size confirmed by TEM. We attribute the fabrication process to the local concentrated field at surface defects and the electrochemical dissolution of the $MoS_2$. The intrinsic electrochemical reaction kinetics permits the ultimate precision for nanopore fabrication. We have observed the step-like features in the ionic current traces, which we attribute to the successive removal of individual atoms. Finally, DNA translocation has been performed to demonstrate the ability such nanopores in detecting molecules. The ECR nanopore fabrication technique presented here offers a well-controlled method to engineer nanopores at single-atom precision and also paves a practical way to scale up the production of 2D nanopores and commercialize nanopore-based technologies.

## ASSOCIATED CONTENT

**Supporting Information**. Experimental methods (setup, CVD $MoS_2$ growth, transfer of CVD $MoS_2$ from sapphire to SiNx membrane, graphene CVD growth, finite element analysis model). Detailed data analysis of ionic current steps presented in **Fig. 3** A reproduced current traces of nanopore formation on $MoS_2$ membrane using ECR showing discrete. **Table S1**. The sequence of cleaving $MoS_2$ unit cells and Mo and S atoms in 21 steps to form the pore, power density spectrum (PSD) noise analysis of ECR fabricated $MoS_2$ nanopore. $\lambda$-DNA translocation trace taken at 300 mV. Analytical model that relates conductance drops to the number and size of the pore (**SI Fig 10.**) **Supporting movie 1s.avi -** displays -the sequence of cleaving $MoS_2$ unit cells and Mo and S atoms in 21 steps to form thepore. Animation based on Cs-STEM image starts from violet, blue, cyan and green to yellow, orange, brown, red and magenta - akin to the visible spectrum sequence. Lifetime of the steps in the sequence is given in the **Table 1.** These times are used as the cues for the animation - the pore formation process that we have recorded is thus shown in real-time. This material is available free of charge *via* the Internet at http://pubs.acs.org.

## ACKNOWLEDGEMENTS

This work was financially supported by European Research Council (grant no. 259398, PorABEL. We would like to thank Prof. Xile Hu for useful discussion. M.L and T.V. acknowledge support by UKF, Grant 17/13, Confined DNA. We thank the Centre Interdisciplinaire de Microscopie Électronique (CIME) at EPFL for access to electron microscopes, and Dr Sorin Lazar of FEI company for assistance with the Cs-STEM TEM imaging. Devices fabrication was partially carried out at the EPFL Center for Micro/Nanotechnology (CMi) and special thanks to Mr. Joffrey Pernollet for assistance in FIB fabrication.

**FIGURES**

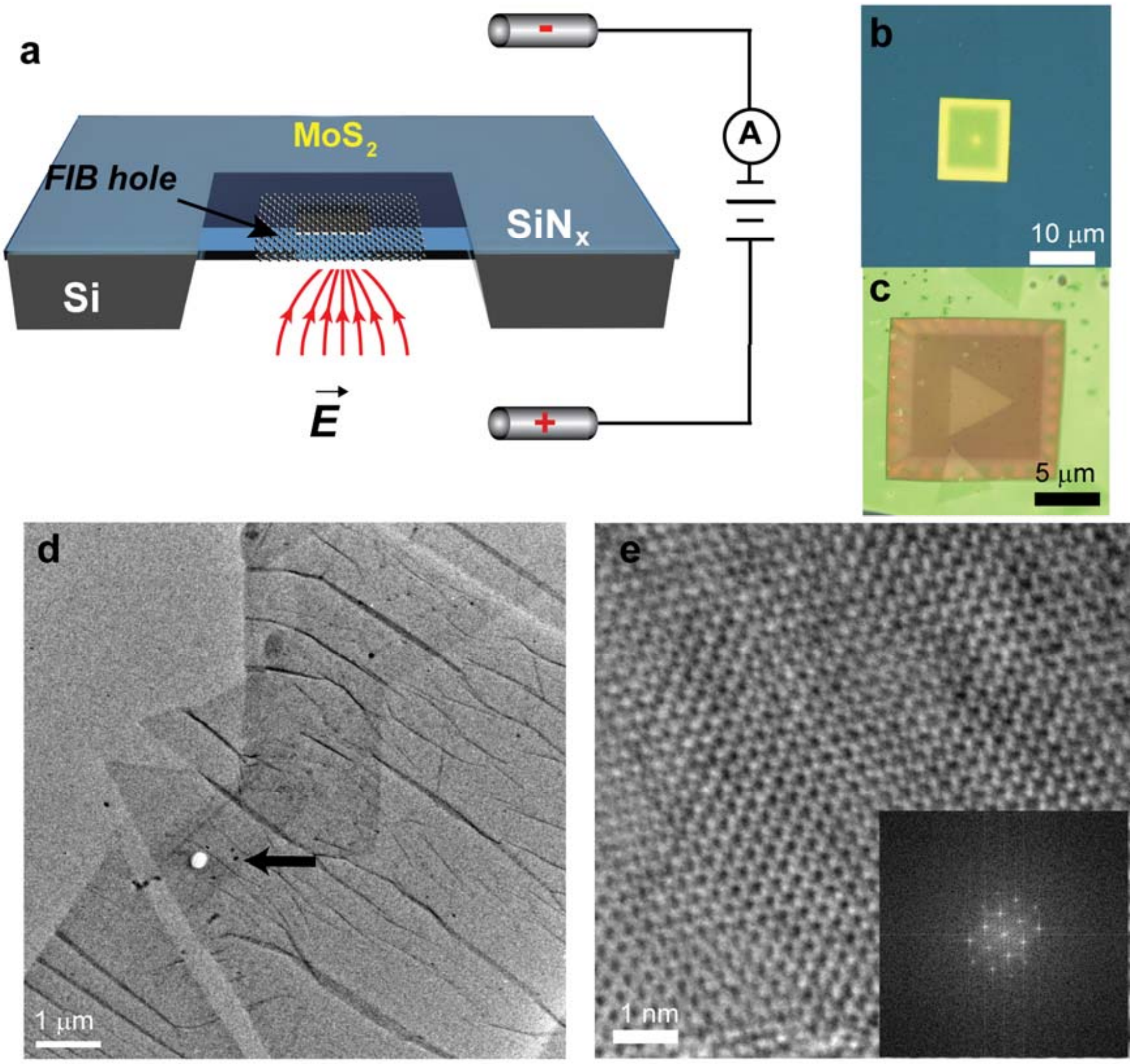

**Fig. 1**. **(a)** Schematic illustration of preparation of a freestanding $MoS_2$ membrane ready for electrochemical formation of a nanopore. In the center of the supporting 20 nm thick $SiN_x$ membrane a single focused ion beam, FIB hole is drilled to suspend a small portion of an intact monolayer $MoS_2$ flake. A single chip is mounted in the flow-cell for typical translocation experiments. A pair of Ag/AgCl electrodes connected to a preamplifier is used to apply transmembrane voltage. **(b)** An optical image of the $SiN_x$ membrane with a FIB drilled hole in the center. **(c)** An optical image of the $SiN_x$ membrane with transferred triangular CVD-grown $MoS_2$ monolayer. **(d)** Low magnification TEM image of transferred CVD-grown $MoS_2$ monolayer covering the FIB hole. The FIB hole is indicated by the black

arrow. **(e)** Conventional high-resolution TEM image of the lattice of $MoS_2$ suspended over the FIB hole. The corresponding diffractogram is shown in the inset.

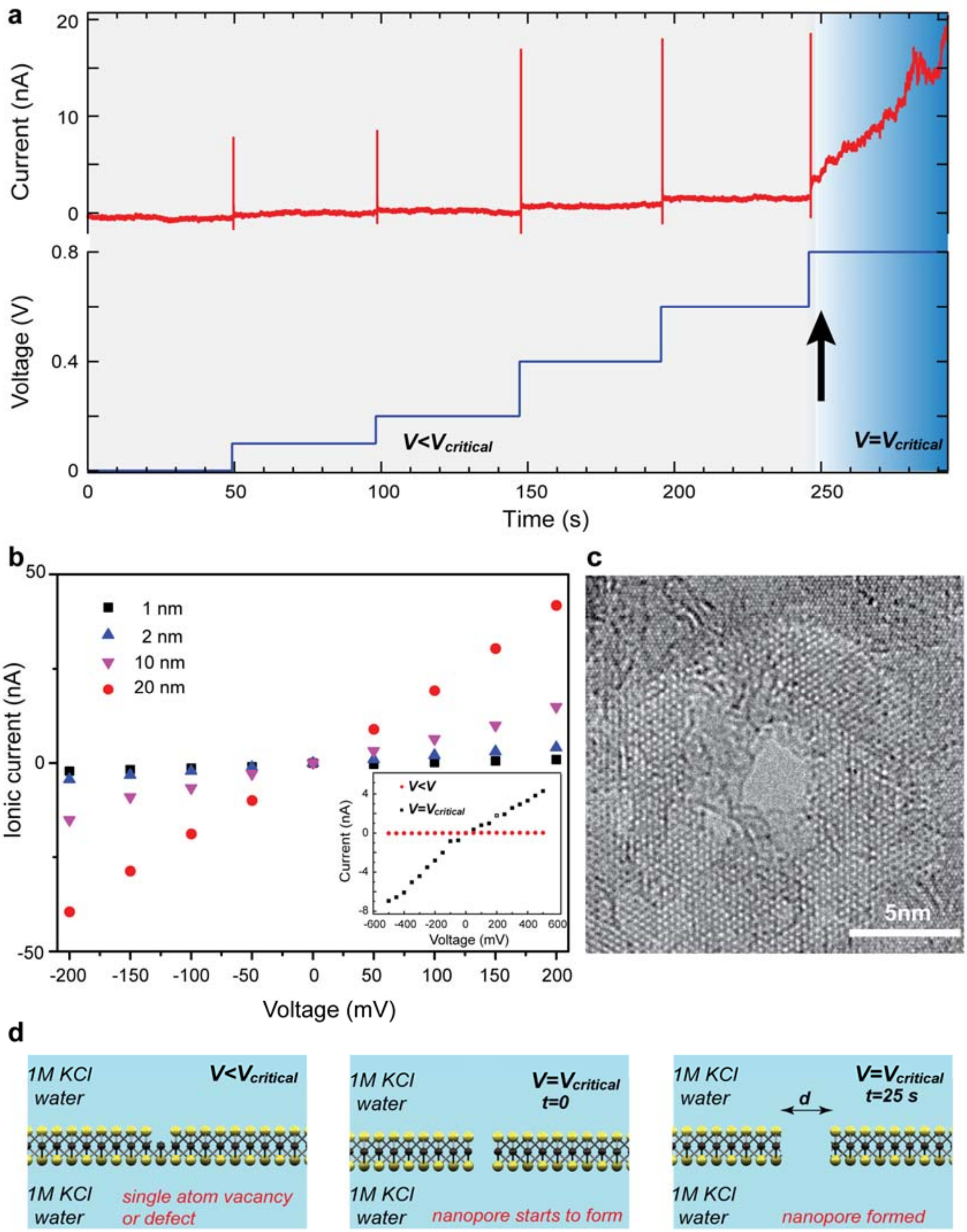

**Fig. 2. (a)** A representative ionic current trace measured for an $MoS_2$ membrane. Voltage is stepped by 100 mV with a 50 s holding, and the leakage current increases in accordance, being steady for a constant voltage. Sharp peaks at each voltage step originate from the capacitance charging. After a critical voltage, 800 mV is applied, the electrochemical reaction, ECR starts (indicated by the black arrow), the current keeps increasing which triggers the feedback control to switch off voltage bias in order to halt the pore growth **(b)**

Current-voltage (IV) characteristic of nanopores ranging in diameter from 1 to 20 nm - all nanopores are created via electrochemical reaction. Inset shows IV characteristics for the system below and at the critical voltage. (**c)** Cs-TEM image taken at 80 keV incident beam energy verifies the nanopore formation and estimated size (3.0 nm) of nanopore created using ECR. (Diameter measured in image ~3 nm). Corresponding current-voltage (IV) characteristic taken after ECR process and prior to Cs-TEM imaging shown in **SI. Fig 4 a.** Larger area (60 nm x60 nm) around ECR created nanopore is shown in **SI.Fig 4 b**. **(d)** Mechanism of ECR based $MoS_2$ nanopore fabrication. A side view of the monolayer $MoS_2$ lattice, emphasizing the lattice having single atom (S) vacancy before ECR $V<V_{critical}$, $MoS_2$ lattice at $V=V_{critical}$ and $MoS_2$ lattice when nanopore is formed.

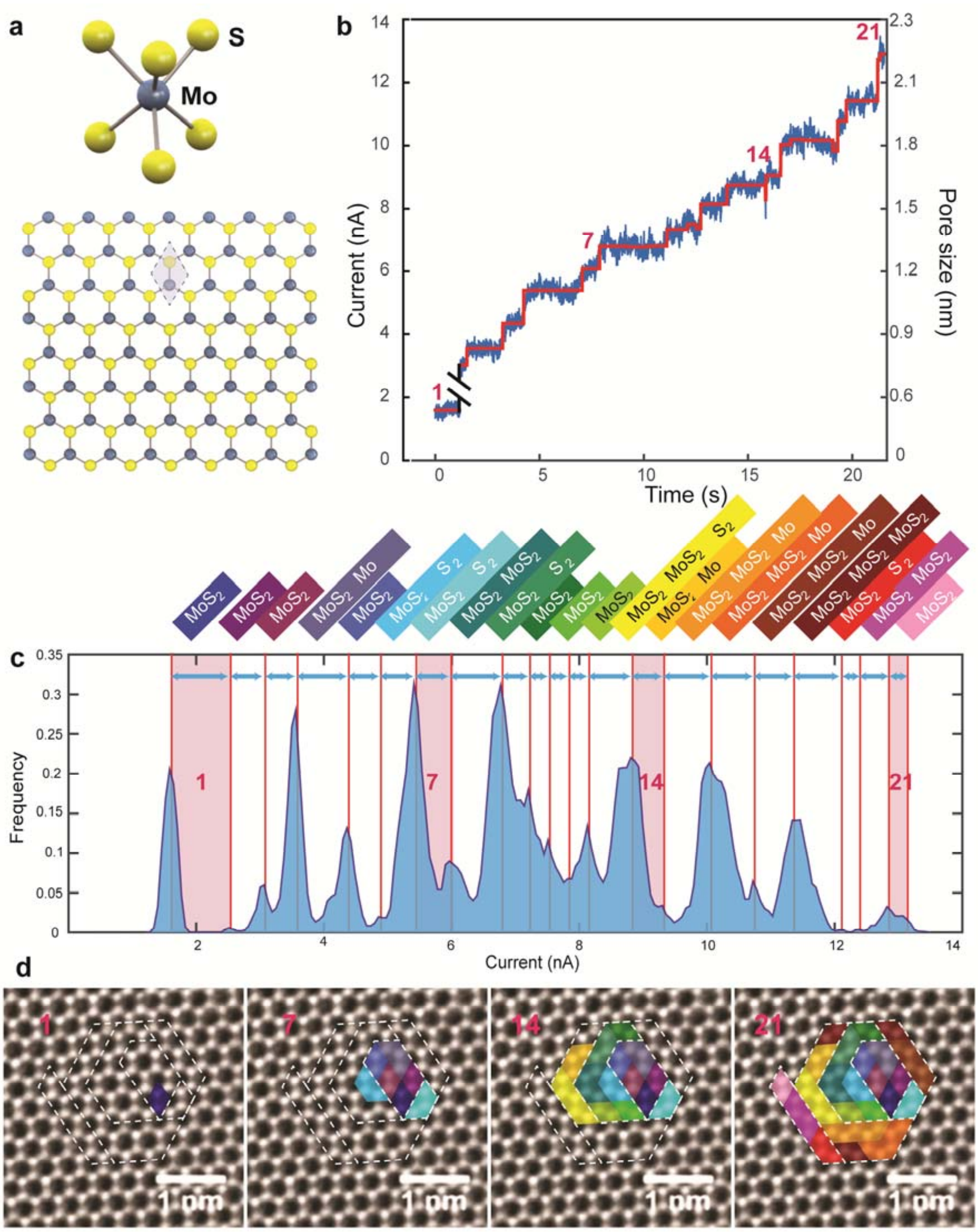

**Fig.** 3. **(a)** A top view of the monolayer $MoS_2$ lattice, the unit cell (parameter a=3.12 Å) is shown in grey[35]. **(b)** Ionic current -steplike features during the nanopore formation in **Fig. 2(a).** A custom Matlab code is used to detect steps in the raw trace[36]. **(c)** Histogram of the trace shown n **(b)** with corresponding color coded atom groups cleaved in each step during the pore formation. **(d)** Illustrative schematic that presents possible outline for nanopore

creation. Polygon removal corresponds to the current histogram trace. Cs-STEM micrograph of suspended single layer $MoS_2$ with superimposed polygons corresponding to atomic groups cleaved in the steps 1,7,14 and 21 during the pore formation. The coloring of atom groups cleaved in each step (**Fig. 3c**) and corresponding area polygons shown in the **Fig. 3d** starts from violet, blue, cyan and green to yellow, orange, brown, red and magenta – analogous to the visible spectrum sequence.

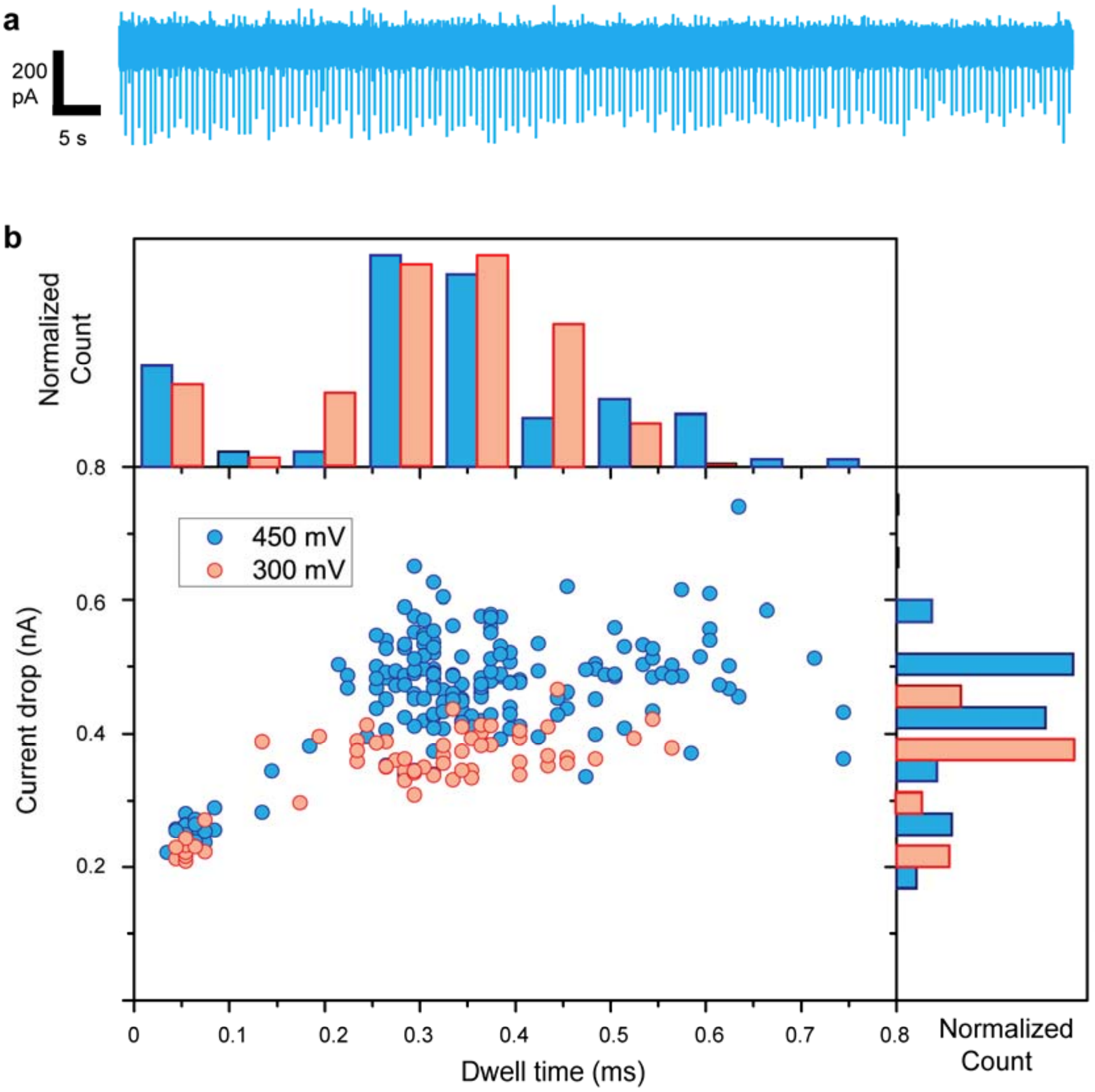

**Fig.** 4. **(a)** A typical trace of pNEB plasmid DNA translocation through an electrochemically etchhed nanopore recorded at 450 mV. The trace is downsampled to 10 kHz for display. **(b)** Scatter plot of events collected at 300 mV and 450 mV bias. Event detection is performed using OpenNanopore[36] Matlab code. Expectedly, the increase in the bias shortens the translocation time and enhances the current drop. Considerably longer term operation of the nanopore device is possible since DNA does not adhere to $MoS_2$, unlike to graphene, as we previously reported[14, 37].

# Supporting Material for:

# Electrochemical reaction in single layer $MoS_2$: nanopores opened atom by atom

J. Feng[1], K. Liu[1], M.Graf[1], M.Lihter[1,2], R. D. Bulushev[1], D. Dumcenco[3], D.T.L. Alexander[4], D. Krasnozhon[3], T. Vuletic[2], A.Kis[3], and A.Radenovic[1]

*[1]Laboratory of Nanoscale Biology, Institute of Bioengineering, School of Engineering, EPFL, 1015 Lausanne, Switzerland*
*[2]Institut za fiziku, Bijenička 46,Zagreb, Croatia*
*[3]Laboratory of Nanoscale Electronics and Structure, Institute of Electrical Engineering , School of Engineering, EPFL, 1015 Lausanne, Switzerland*
*[4]Centre Interdisciplinaire de Microscopie Électronique (CIME), 1015 Lausanne, Switzerland*

# Table of contents

# Experimental Methods

## Setup

The $MoS_2$ membranes are  prepared using the previously reported procedure[1]. Briefly, 20 nm thick supporting $SiN_x$ membranes are manufactured using anisotropic KOH etching to obtain 10 μm × 10 μm to 50 μm × 50 μm membranes, with size depending on the size of the backside opening. Focused ion beam (FIB) is used to drill a 50 − 300 nm opening on that membrane. CVD-grown $MoS_2$ flakes were transferred from sapphire substrates using $MoS_2$ transfer stage in a manner similar to the widely used graphene transfer method and suspended on FIB opening[2,3]. Membranes are first imaged in the TEM with low magnification in order to check suspended $MoS_2$ flakes on FIB opening.

For the nanopore fabrication experiments, after mounting in the polymethylmethacrylate (PMMA) chamber, the chips were wetted with $H_2O$:ethanol (v:v, 1:1) for at least 20 min. 1 M KCl solution buffered with 10mM Tris-HCl and 1mM EDTA at pH 8.0 was injected in the chamber. A pair of chlorinated Ag/AgCl electrodes was employed to apply the transmembrane voltage and the current between the two electrodes was measured by a FEMTO DLPCA-200 amplifier (FEMTO® Messtechnik GmbH)A low voltage (100 mV) was applied to check the current leakage of the membrane. If the leakage current was below 1 nA, we stepped-up the voltage bias in 100 mV steps (25 s for each step). At a critical voltage we observed the current starting to immediately increase above the leakage level. We use a FPGA card and custom-made LabView software for applying the voltage. The critical voltage was automatically shut-down by a feedback control implemented in LabView program as soon as the desirable conductance was reached. Nanopores were further imaged using a JEOL 2200FS high-resolution transmission electron microscope (HR-TEM). Scanning TEM (STEM) energy dispersive X-ray spectroscopy (EDX) mapping was performed on a ChemiSTEM-equipped FEI Tecnai Osiris transmission electron microscope (TEM). Aberration-corrected TEM micrographs were taken on a FEI Titan Themis 60-300 at 80 keV.

Current–voltage, IV characteristic and DNA translocation were recorded on an Axopatch 200B patch clamp amplifier (Molecular Devices, Inc. Sunnyvale, CA). DNA samples (pNEB193, plasmid 2.7 k bp, New England; λ-DNA, 48 k bp, New England) were diluted by mixing 10 μL of λ-DNA or pNEB stock solution with 490 μL 1 M KCl buffer. We use a NI PXI-4461 card for data digitalization and custom-made LabView software for data acquisition using Axopatch 200B. The sampling rate is 100 kHz and a built-in low-pass filter at 10 kHz is used. Data analysis enabling event detection is performed offline using a custom open source Matlab code, named OpenNanopore[4] (http://lben.epfl.ch/page-79460-en.html).

**CVD $MoS_2$ growth**

Monolayer $MoS_2$ has been grown by chemical vapor deposition (CVD) on c-plane sapphire. After consecutive cleaning by acetone/isopropanol/DI-water the substrates were annealed for 1h at 1000 ℃ in air. After that, they were placed face-down above a crucible containing ~5 mg $MoO_3$ (≥ 99.998% Alfa Aesar) and loaded into a furnace with a 32 mm outer diameter quartz tube. CVD growth was performed at atmospheric pressure using ultrahigh-purity argon as the carrier gas. A second crucible containing 350 mg of sulfur (≥ 99.99% purity, Sigma Aldrich) was located upstream from the growth substrates. More details are available in [3].

**CVD $MoS_2$ transfer from sapphire to $SiN_x$ membrane**

Monolayer $MoS_2$ grown on sapphire substrate (12 mm by 12 mm) is coated by A8 PMMA (495) and baked at 180 ℃. We use a diamond scriber to cut it into 4 pieces. Each piece is immersed into 30%w KOH at 85-90 ℃ for the detachment. It is advisory to use capillary force in the interface between polymer and sapphire to facilitate the detachment and reduce the etching time in the KOH. The detached polymer film was repeatedly in DI water. Lastly, the "fishing" method of graphene transfer can be used to transfer CVD $MoS_2$ to the target $SiN_x$ membrane.

**CVD graphene growth**

Large-area graphene films are grown on copper foils. The growth takes place under the flow of a methane / argon / hydrogen reaction gas mixture at a temperature of 1000 ℃. At the end of the growth, the temperature is rapidly decreased and the gas flow turned off. The copper foils are then coated by PMMA and the copper etched away, resulting in a cm-scale graphene film ready to be transferred on the chips with membranes.

## Finite element analysis model

To estimate the potential drop in a defect in a $MoS_2$ membrane a finite element analysis was performed using COMSOL Multiphysics 4.4b. A coupled set of the Poisson-Nernst-Planck equations was solved in a 3D geometry with axial symmetry. In the modeled configuration cis and trans chambers were connected by a 0.3 nm pore in a 0.7 nm thick membrane suspended on a 50 nm wide and 20 nm thick hole. A 0.3 nm diameter defect can correspond to the absence of a unit cell of $MoS_2$. In the model, the applied potential was set to 800 mV and salt concentration was 1 M KCl. The minimal mesh size used was less than 0.2 Å.

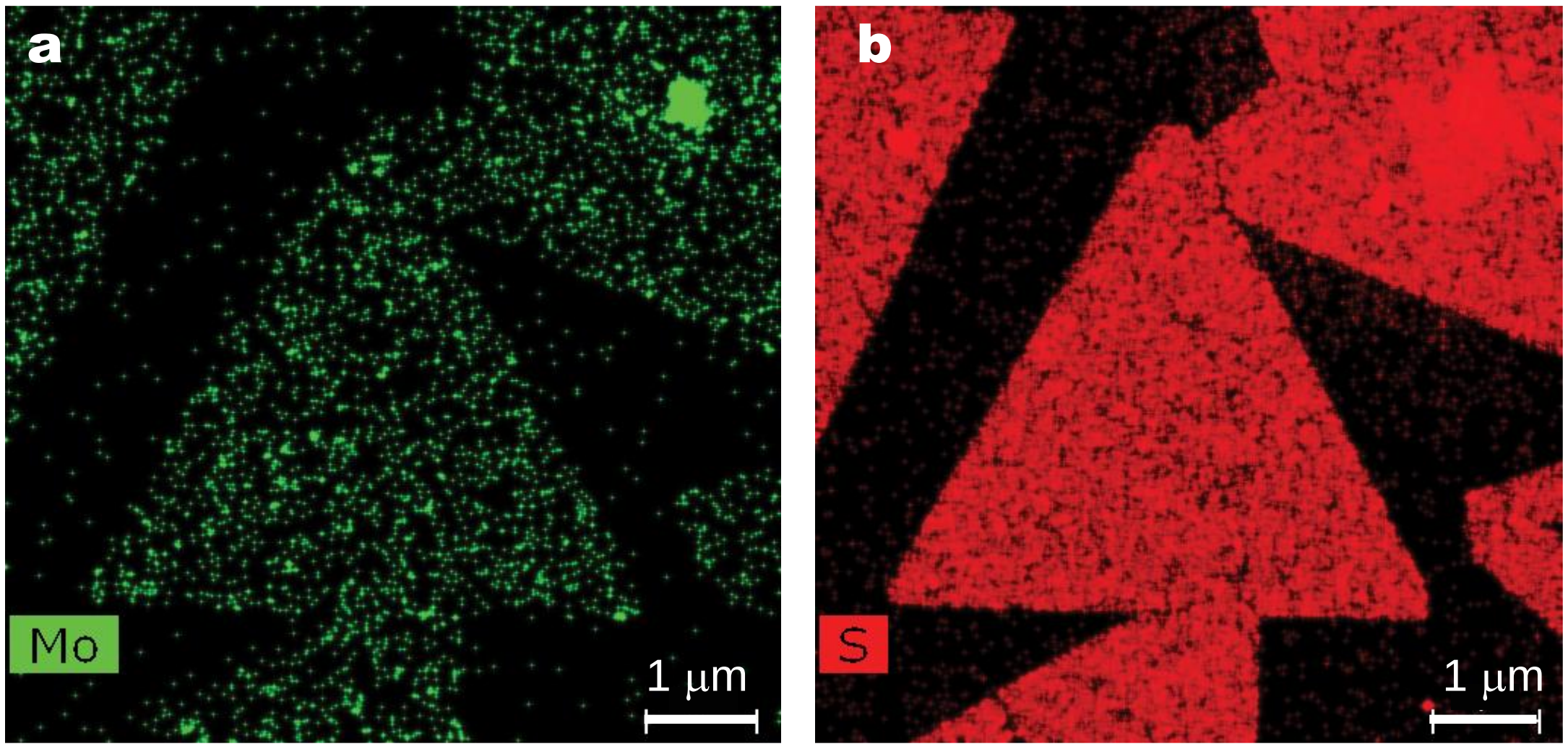

**SI Fig.1** EDX mapping of Mo and S elements in the monolayer $MoS_2$ film composed of triangular single-crystal domains transferred on the supporting $SiN_x$ membrane. FEI Tecnai Osiris TEM is operated in the STEM mode at 200 kV to achieve high speed and high sensitivity EDX measurements. To unambiguously decouple S from Mo Electron Energy Loss Spectroscopy (EELS) analysis of the samples would be required.

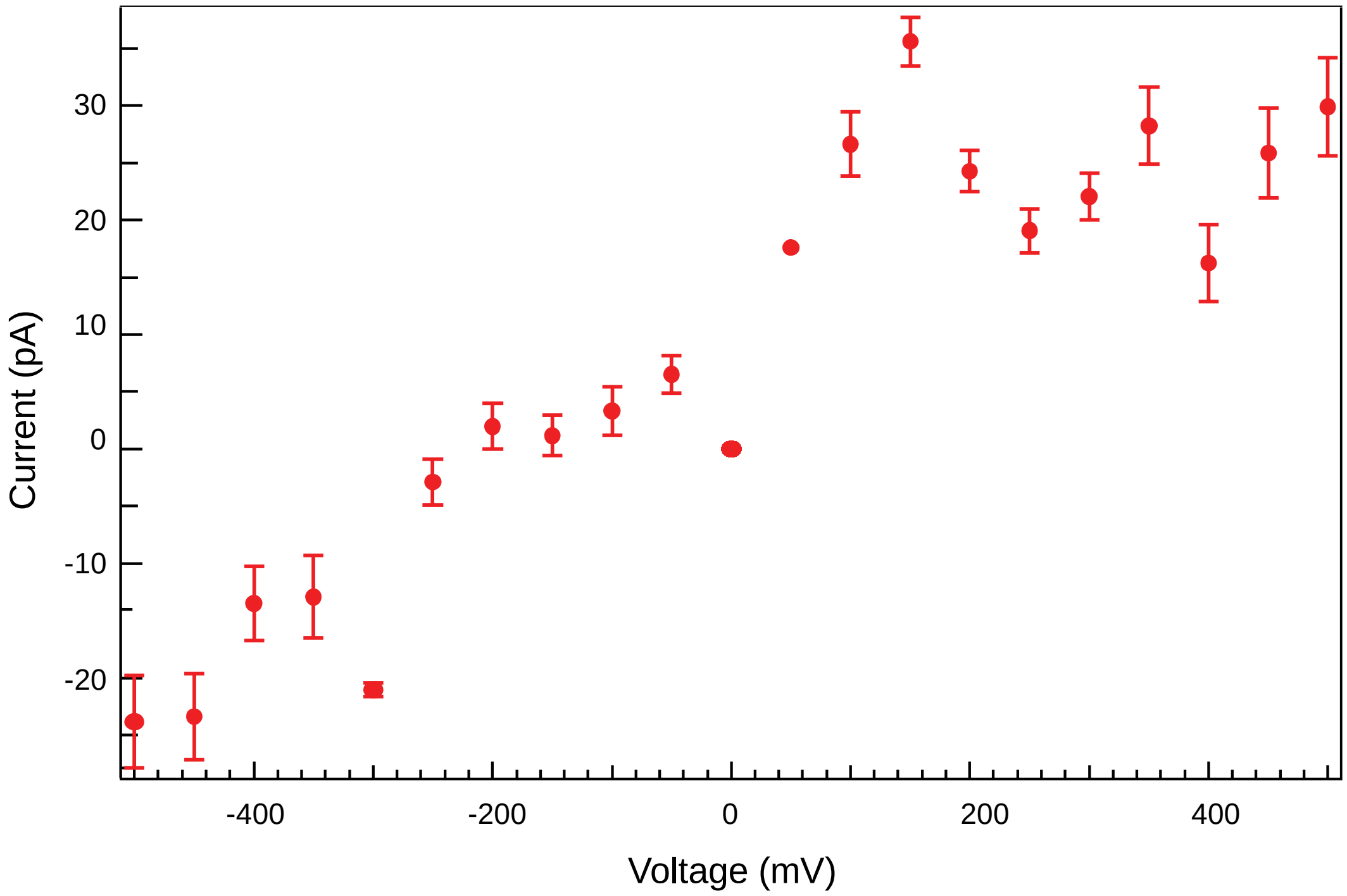

**SI Fig.2** Leakage current-voltage (IV) characteristic of an intact $MoS_2$ membrane, for voltages below the critical voltage of 800 mV required for ECR. The leakage current depends on the number of the membrane defects. More defects leads to higher current.

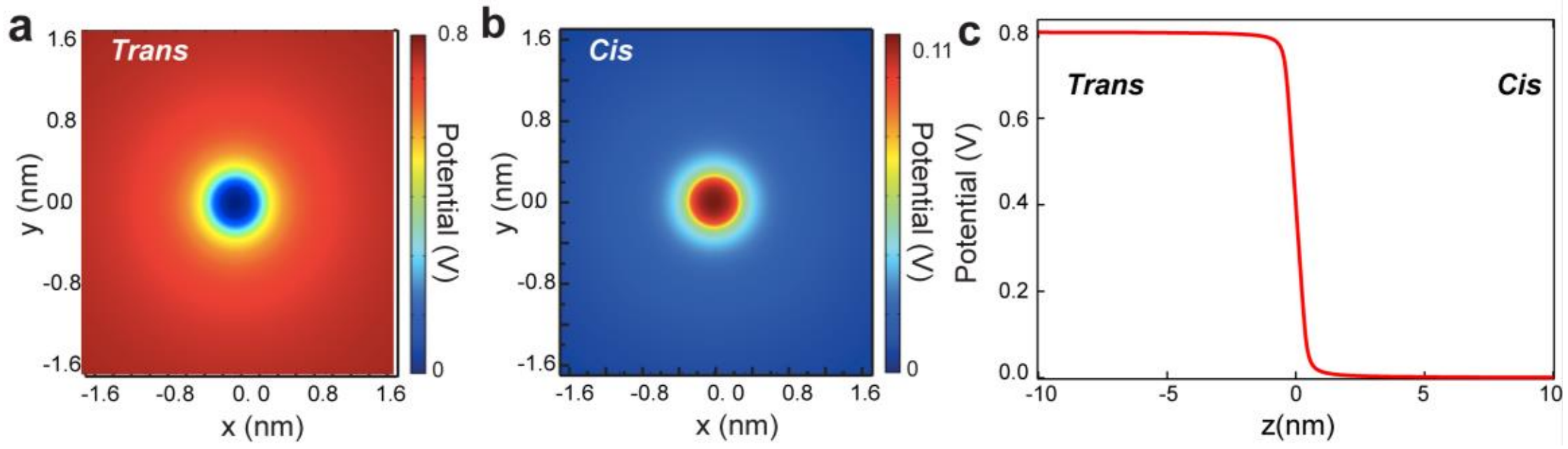

**SI Fig.3. Simulations of the electric potential distribution for the nanopore in two dimensions for a just formed pore having a diameter of 0.3 nm. (a)** Electric potential distribution in the *trans* chamber in the immediate vicinity of the membrane surface and **(b)** in the *cis* chamber. **(c)** Electric potential distribution as a function of the distance from the pore. The applied potential was set to 800 mV and salt concentration was 1 M KCl.

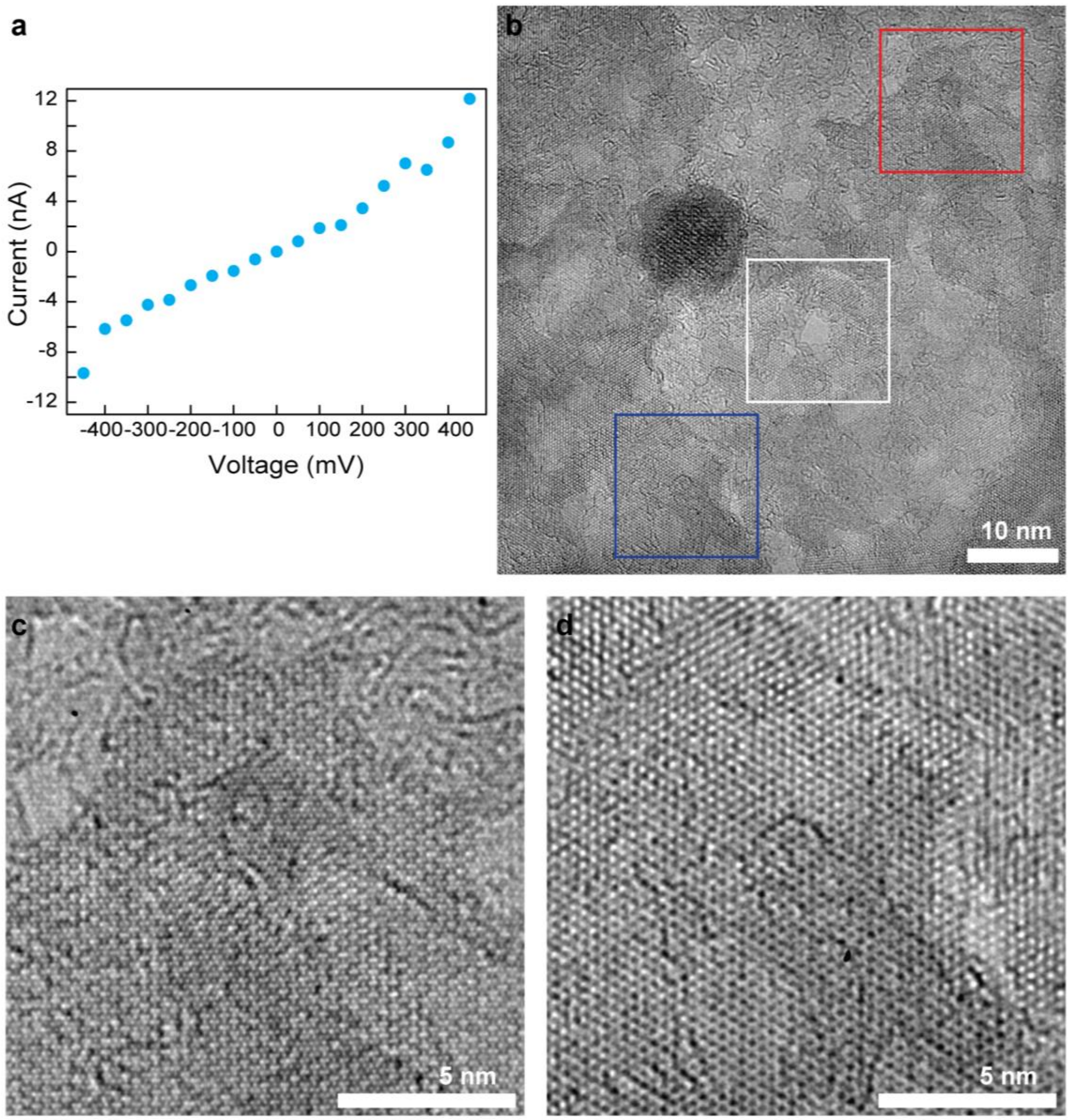

**SI Fig.4. (a)** Current-voltage (IV) characteristic of nanopore created via electrochemical reaction having conductance of ~ 22.8 nS in 1M KCl which corresponds to nanopore having ~3.0 nm diameter. **(b)** Large field of view area (60 nm x 60 nm) Cs-TEM image around the ECR-created nanopore in the middle of the white square, which corresponds to the zoomed region shown in **Fig.2.c. (c)** and **(d)** show random (15nm x15 nm ) zooms in the regions indicated by red **(c)** and blue **(d)** squares in **(b)**.

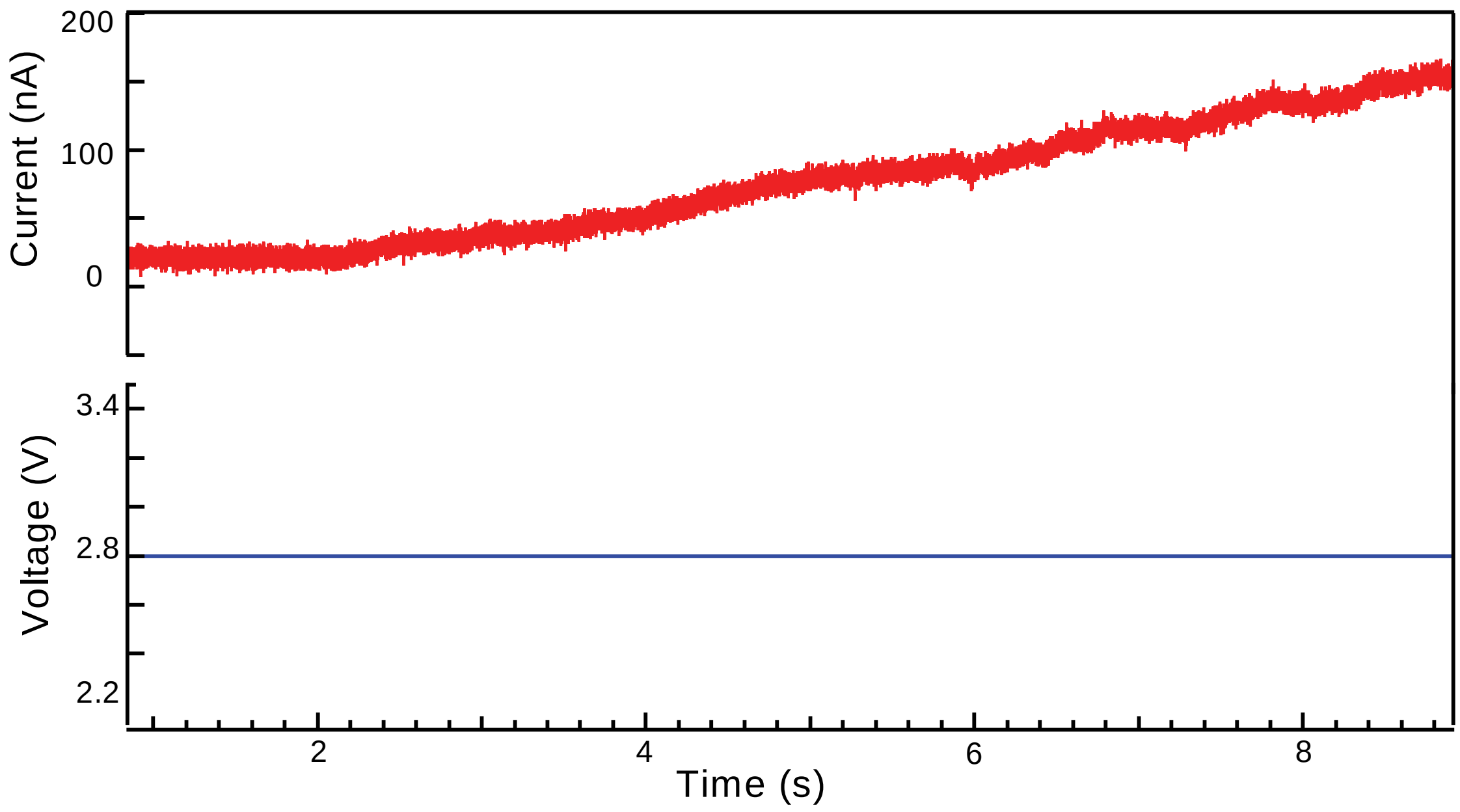

**SI Fig.5** A typical current trace of nanopore formation on graphene membrane using ECR. A much higher transmembrane voltage, 2.8 V has to be applied to graphene to create a nanopore in graphene.

## 6. Detailed data analysis of ionic current steps presented in Fig. 3.

All analysis were implemented in Matlab R2014b. The raw signal was down-sampled to 5 kHz and then filtered using the edge-preserving Chung-Kennedy (CK) filter[5] (**Fig. S5a**). The pore formation in 21 steps, presented in the journal article on Fig. 3 can be understood in the following way. The growth of the nanopore is due to sequential cleaving of unit cells from $MoS_2$ lattice. The final pore area is 2.9 $nm^2$, which corresponds to 34 unit cells. Increments in the effective pore size $\Delta A$ are normalized by unit cell size $u = 0.0864\ nm^2$. We round the obtained number $\Delta N = \Delta A/u$ to the nearest integer, integer +1/3 or integer +2/3 to get $\Delta X$, the number of $MoS_2$ unit cells cleaved during the pore formation process. We assume that 1/3 corresponds to a $S_2$ group and 2/3 to a Mo atom, corresponding to the partial cleaving of a unit cell. It should be noted that the two S atoms in $MoS_2$ are stacked vertically and their combined surface area is smaller than that for Mo (which has about 50% larger radius).

The sequence of cleaving $MoS_2$ unit cells and Mo and S atoms in 21 steps to form the pore is given in the **Table S1.** In order to depict the sequence of the pore formation, the coloring of the lines in the **Table S1** and polygons in the animation based on HRTEM image starts from violet, blue, cyan and green to yellow, orange, brown, red and magenta – akin to the visible spectrum sequence (see ***Supporting Movie 1.avi***). Lifetime of the steps in the sequence is given in the **Table 1**. These times are used as the cues for the animation - the pore formation process that we have recorded is thus shown in real-time. Notably, initially irregular pore gradually becomes more symmetrical. White dashed line is a visual aid to denote the progress of pore growth. The atom groups have been selected in the manner to minimize the number of dangling bonds at the edge of the pore. The pore formation sequence is not unique, however, the dangling bond constraint significantly reduces the number of pore formation scenarios and induces more symmetrical pore shape.

| I step [nA] | lifetime [s] | D [nm] | A [nm²] | N | ΔA [nm²] | ΔN | ΔX: | atoms/groups cleaved | | |
|---|---|---|---|---|---|---|---|---|---|---|
| 0,926 | 0,1 | 0,36 | 0,10 | 1,16 | 0,100 | 1,16 | 1 | | $MoS_2$ | |
| 0,5556 | 0,2 | 0,47 | 0,18 | 2,04 | 0,076 | 0,88 | 1 | | $MoS_2$ | |
| 0,4939 | 1,7 | 0,57 | 0,25 | 2,92 | 0,076 | 0,88 | 1 | | $MoS_2$ | |
| 0,8025 | 1,0 | 0,71 | 0,39 | 4,53 | 0,139 | 1,60 | 1 2/3 | | $MoS_2$ | Mo |
| 0,4939 | 0,1 | 0,79 | 0,49 | 5,62 | 0,094 | 1,09 | 1 | | $MoS_2$ | |
| 0,5556 | 2,7 | 0,87 | 0,60 | 6,93 | 0,114 | 1,31 | 1 1/3 | | $MoS_2$ | $S_2$ |
| 0,5556 | 0,8 | 0,96 | 0,72 | 8,34 | 0,122 | 1,41 | 1 1/3 | | $MoS_2$ | $S_2$ |
| 0,8025 | 3,2 | 1,08 | 0,91 | 10,53 | 0,189 | 2,19 | 2 | $MoS_2$ | $MoS_2$ | |
| 0,4322 | 1,5 | 1,14 | 1,02 | 11,79 | 0,109 | 1,26 | 1 1/3 | | $MoS_2$ | $S_2$ |
| 0,3087 | 0,1 | 1,18 | 1,10 | 12,72 | 0,080 | 0,93 | 1 | | $MoS_2$ | |
| 0,3087 | 0,1 | 1,23 | 1,18 | 13,67 | 0,083 | 0,96 | 1 | | $MoS_2$ | |
| 0,3087 | 1,1 | 1,27 | 1,27 | 14,66 | 0,085 | 0,98 | 1 | | $MoS_2$ | |
| 0,6791 | 1,8 | 1,36 | 1,46 | 16,92 | 0,195 | 2,26 | 2 1/3 | $MoS_2$ | $MoS_2$ | $S_2$ |
| 0,4939 | 0,7 | 1,43 | 1,61 | 18,64 | 0,149 | 1,72 | 1 2/3 | | $MoS_2$ | Mo |
| 0,7408 | 2,7 | 1,53 | 1,84 | 21,35 | 0,234 | 2,71 | 2 2/3 | $MoS_2$ | $MoS_2$ | Mo |
| 0,6791 | 0,4 | 1,62 | 2,07 | 23,96 | 0,226 | 2,62 | 2 2/3 | $MoS_2$ | $MoS_2$ | Mo |
| 0,6173 | 1,4 | 1,71 | 2,29 | 26,45 | 0,215 | 2,49 | 2 2/3 | $MoS_2$ | $MoS_2$ | Mo |
| 0,7408 | 0,1 | 1,80 | 2,55 | 29,57 | 0,270 | 3,12 | 3 | $MoS_2$ | $MoS_2$ | $MoS_2$ |
| 0,3087 | 0,1 | 1,84 | 2,67 | 30,91 | 0,116 | 1,34 | 1 1/3 | | $MoS_2$ | $S_2$ |
| 0,4321 | 0,1 | 1,90 | 2,84 | 32,84 | 0,166 | 1,92 | 2 | $MoS_2$ | $MoS_2$ | |
| 0,2469 | 0,5 | 1,93 | 2,93 | 33,96 | 0,097 | 1,12 | 1 | | $MoS_2$ | |
| **Pore:** | | **Diameter:** | **size nm²** | **cells** | | | | | | |
| | | 1,9 nm | 2,9 | 34,0 | | | 34 | | | |

**Table 1. The sequence of cleaving $MoS_2$ unit cells and Mo and S atoms in 21 steps to form the pore.**

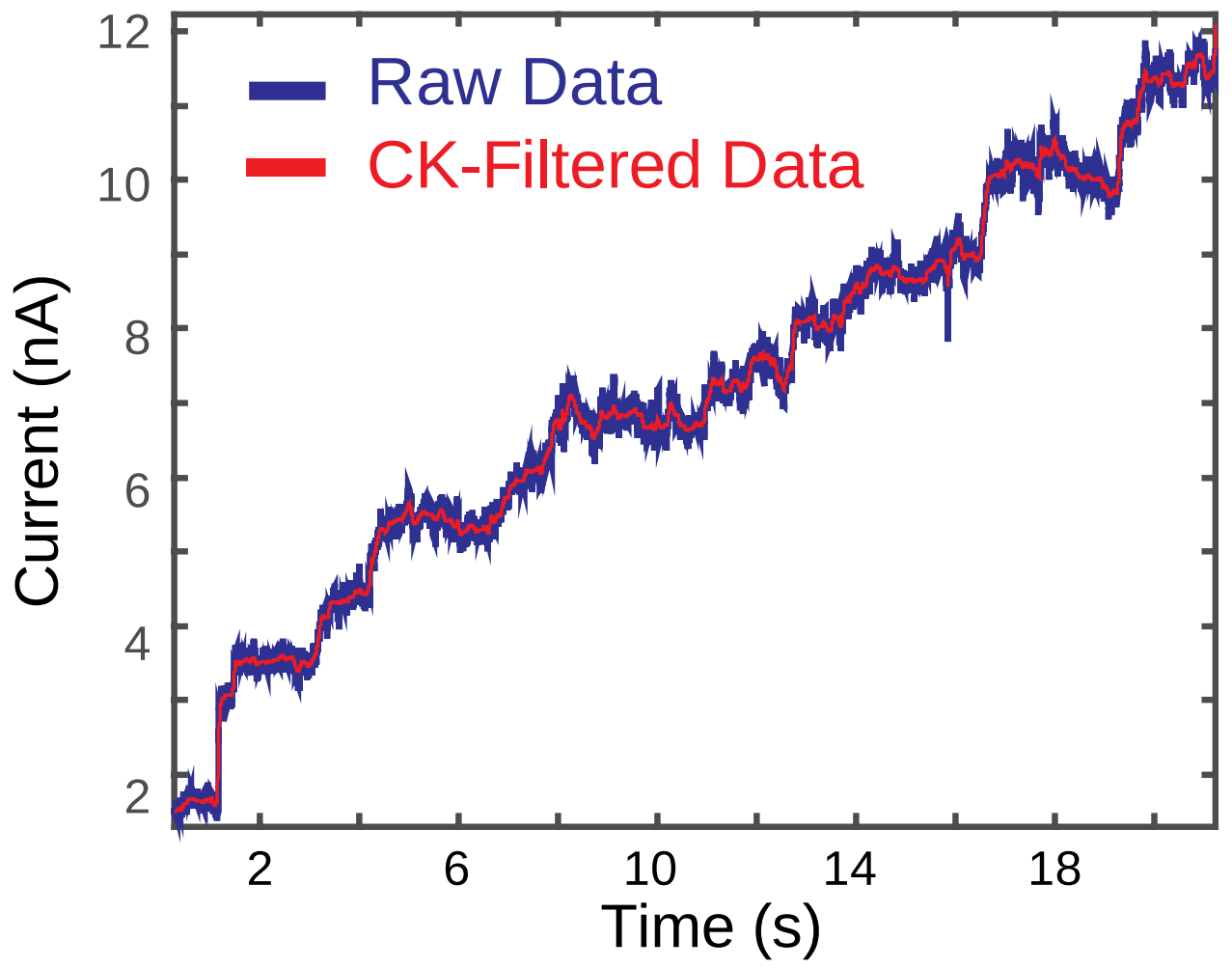

**SI Fig.6** Ionic current signal, obtained during ECR reaction and pore formation. The raw signal was down-sampled to 5 kHz and filtered with the edge-preserving Chung-Kennedy (CK) filter.

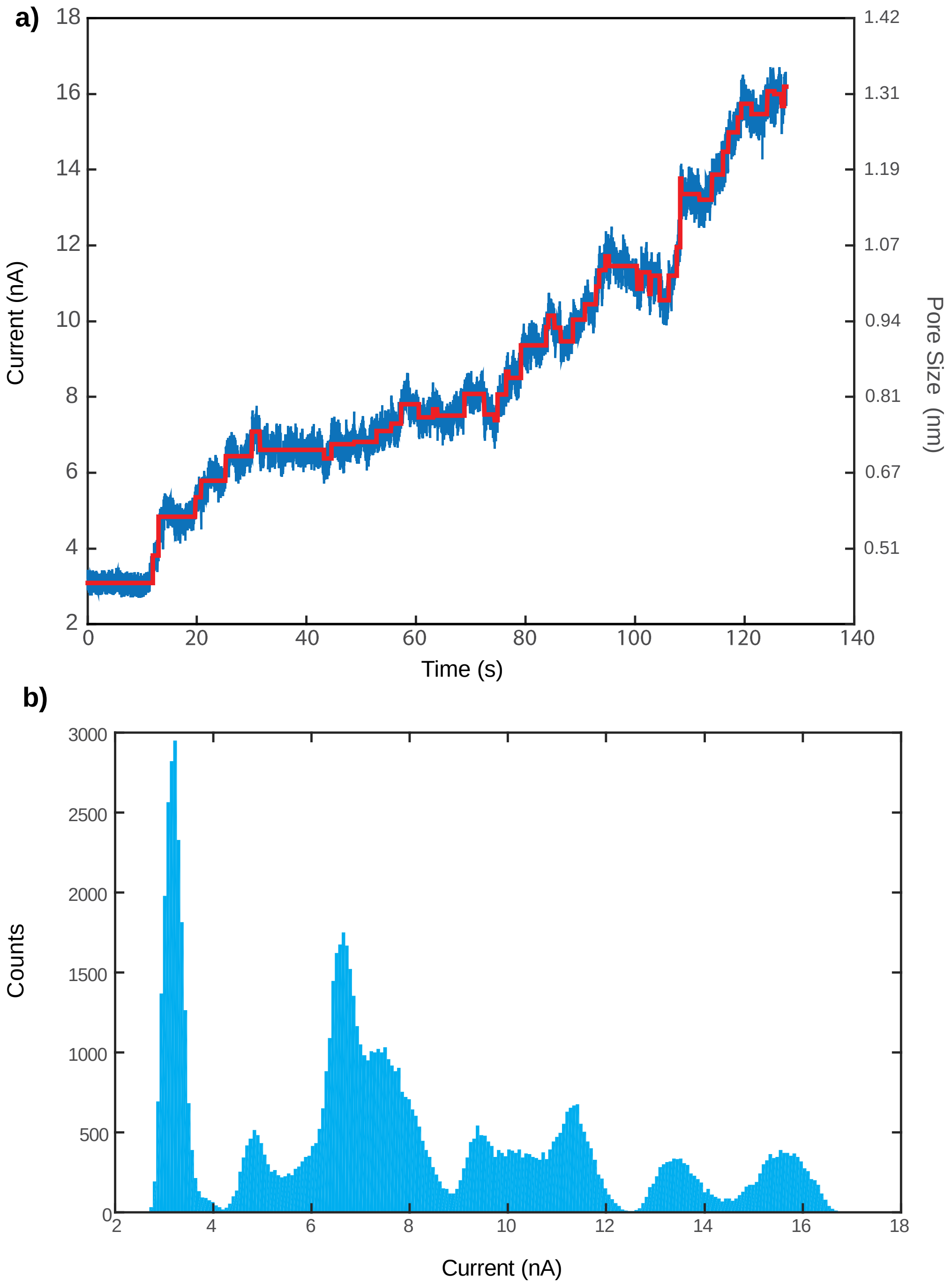

**SI Fig.7 (a)** A reproduced current trace of nanopore formation on $MoS_2$ membrane using ECR method showing discrete steps at critical potential of 2 V **(b)** Corresponding histogram.

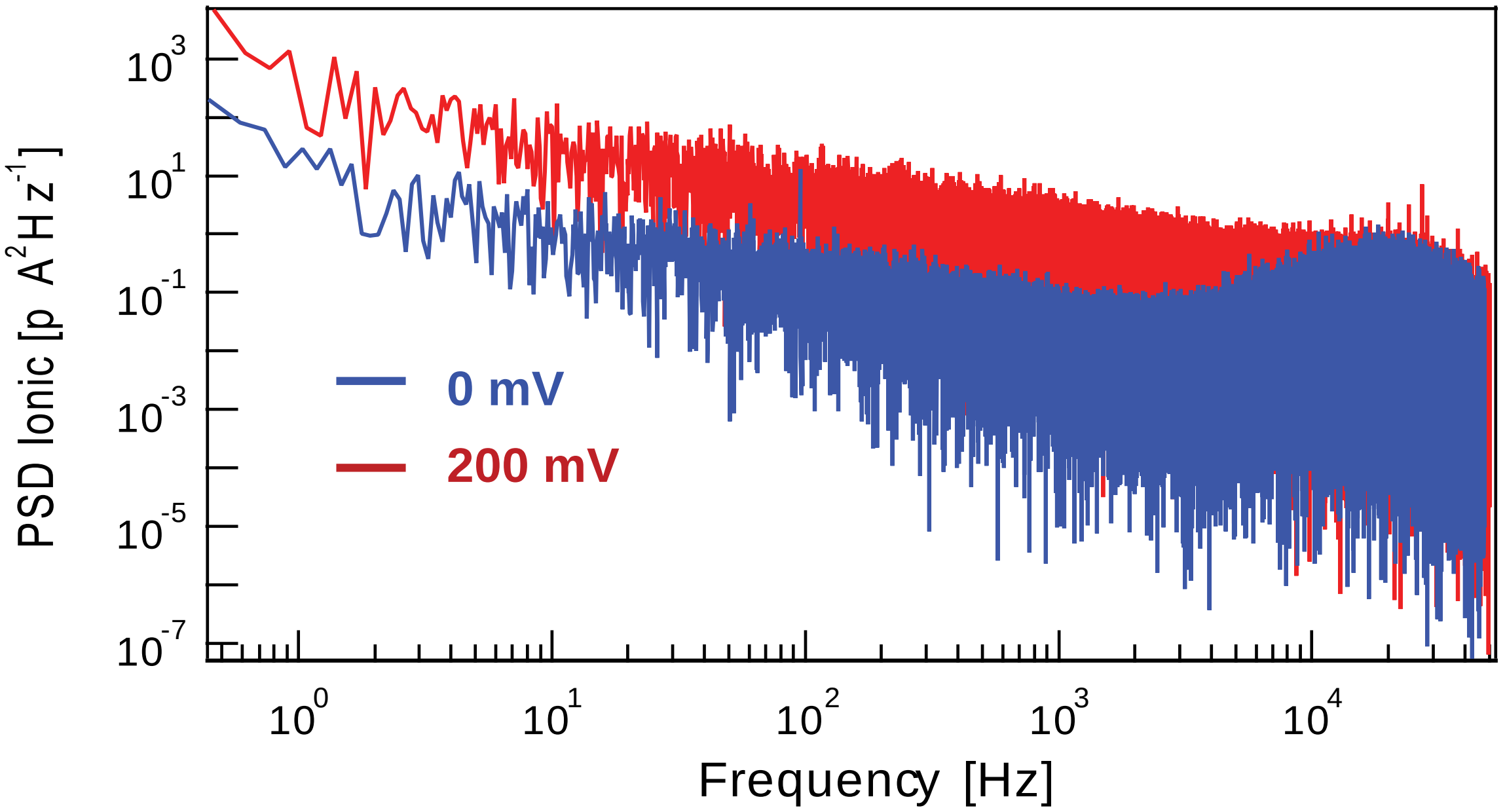

**SI Fig.8** Power density spectrum (PSD) noise analysis of an ECR fabricated $MoS_2$ nanopore at the transmembrane voltages of 0 mV (blue) and 200 mV (red), respectively. A short fragment at each voltage of blank ionic current trace is chosen for such an analysis.

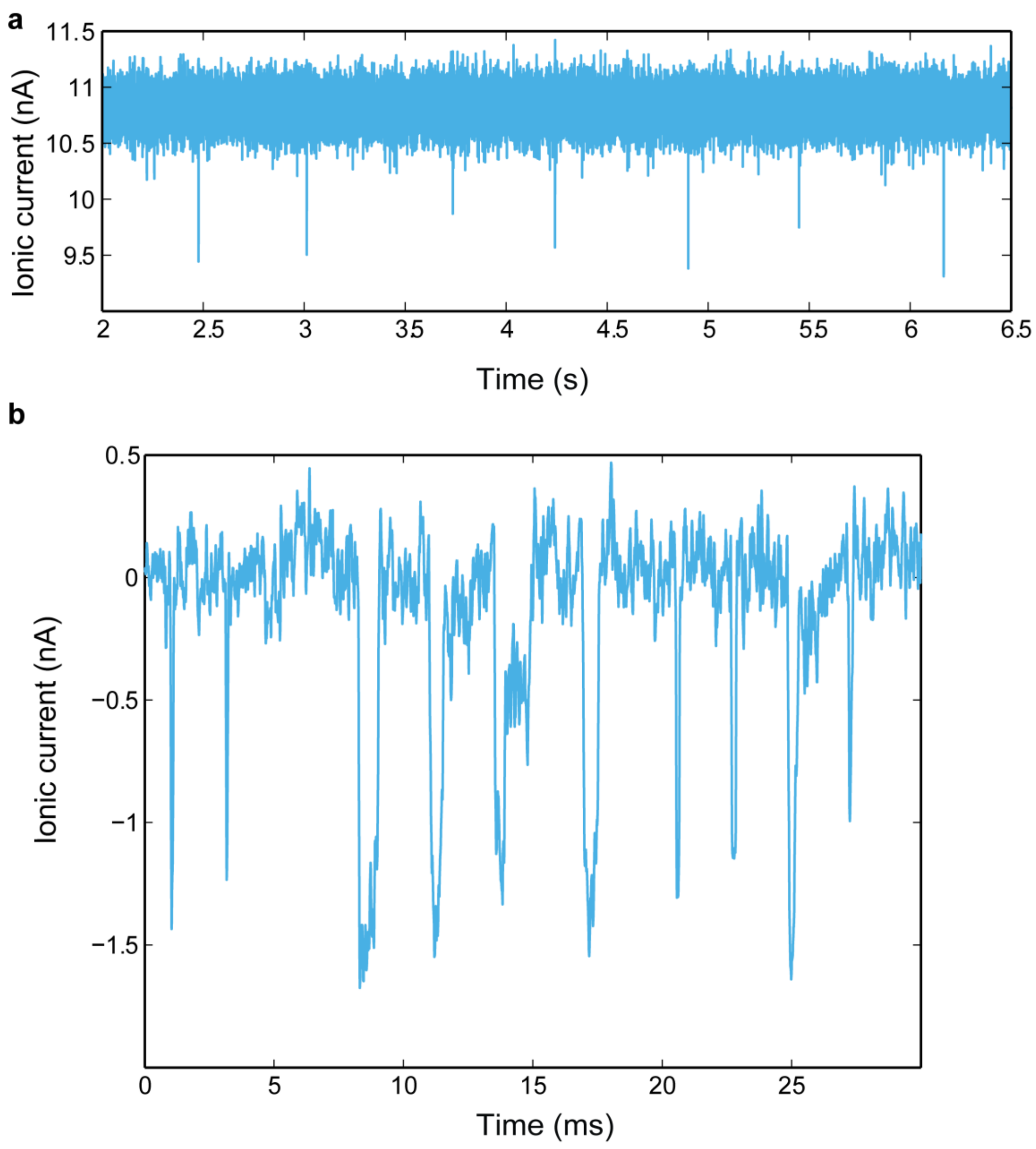

**SI Fig.9. (a)** Long trace showing the λ-DNA translocations through a 4.3 nm ECR fabricated $MoS_2$ nanopore recorded in-situ right after pore formation at 300 mV. **(b)** Concatenated λ-DNA translocation events.

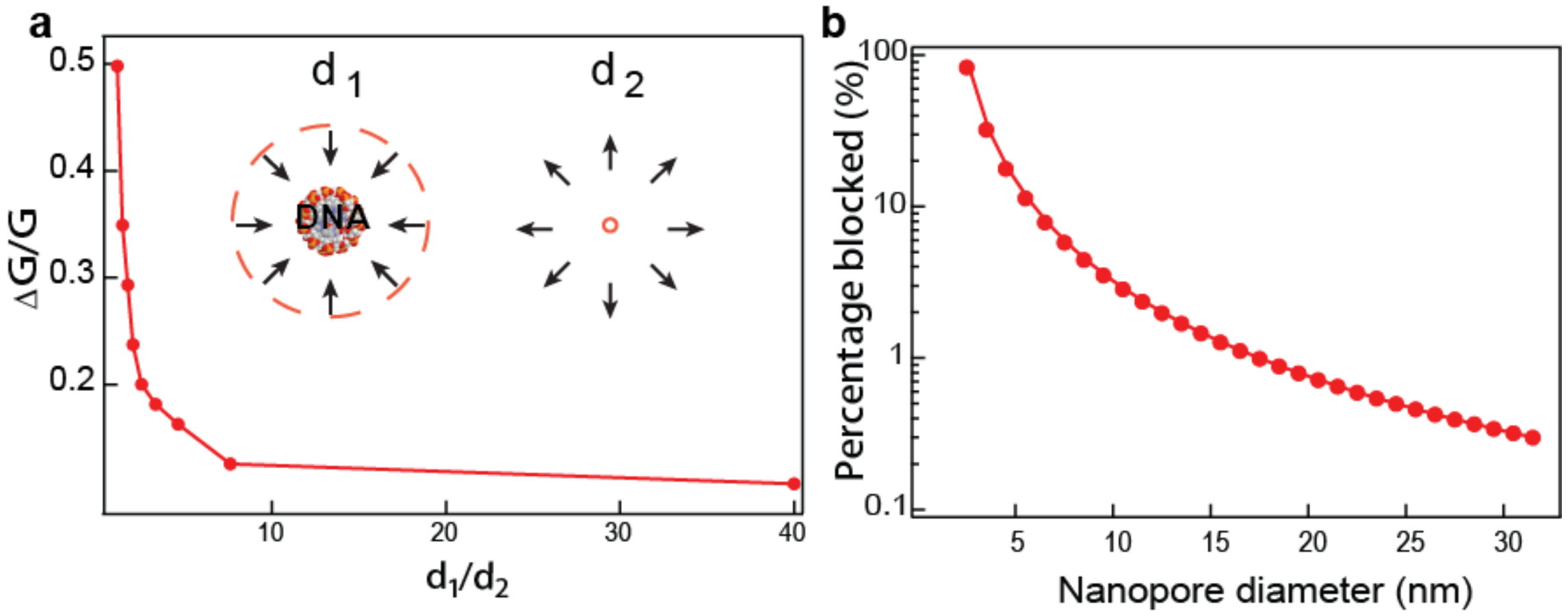

**SI Fig.10**. **(a)** Simplistic analytical model that relates normalized conductance drops to the ratio of the sizes of the 2 pores. Initial nanopore diameters are set to $d_1$=4 and $d_2$= 0.1 nm. We varied the sizes of the both pores while keeping the total conductance fixed. **(b)** Percentage of the blocked ionic current as a function of nanopore diameter. More rigorous model is provided by Garaj et al. [6]. Surprisingly our simplistic model agrees well with Garaj et al. [6]